\documentstyle[prl,aps,preprint,epsfig]{revtex}

\tightenlines

\begin{document}

\date{\today}
\draft

\title{
A Transport Analysis of the BEEM Spectroscopy
of Au/Si Schottky Barriers} 

\author{
U. Hohenester$^{a,b}$, P. Kocevar $^{b}$, P. de Andres
$^{c}$, and F. Flores $^{d}$}

\address{ 
$^{a}$E-mail: {\small uxh@physik.kfunigraz.ac.at}, 
FAX: {\small +43 316 380 9820}\\
$^{b}$Inst. f. Theoret. Physik, Karl--Franzens--Univ., A--8010 Graz,
Austria\\
  $^{c}$Inst. de Ciencia de Materiales (CSIC), E--28049 Madrid, Spain\\
  $^{d}$Dep. de Fisica de la Mat. Cond., Univ. Auton.,
    E--28049 Madrid, Spain
}

\maketitle

\begin{abstract}
A systematic transport study of the ballistic electron emission
microscopy ({\sc beem}) of Au/Si(100) and
Au/Si(111) Schottky barriers for different thicknesses of the metal layer
and different temperatures is presented. It is shown that the existing
experimental data are compatible with a recently predicted
bandstructure--induced non--forward electron propagation through the
Au(111) layer.
\end{abstract}

REF.: Phys. Stat. Sol. (b), submitted. 

\section{Introduction}

Ballistic-electron emission microscopy ({\sc beem}) has proven to
be an extremely efficient tool to study metal-semiconductor interfaces
on a nanometer scale \cite{Kaiser/Bell:88,Prietsch:95}.
{\sc beem} is a technique based on scanning tunneling microscopy ({\sc
stm}) where the {\sc stm} tip is used as an electron source for
a highly space--resolved injection into the metal layer. By collecting the
current which passes
from the metal into the semiconductor as a function of tip position and
tip bias, information about the local Schottky barrier height and the
hot-electron transport properties can be obtained on a nanometer
scale.

Although the Au/Si interfaces have been among the systems most
thoroughly studied with the {\sc beem} technique, they also turned out
to be among the most controversial ones. In particular, the very
similar {\sc beem} spectra for Au/Si(100) and Au/Si(111) systems
\cite{Schowalter/Lee:91} have been a matter of intensive debate
\cite{Prietsch:95}.  A number of theoretical models has been brought up
to account for these experimental findings. While these models have
been sucessful in the interpretation of particular systems, only few
attempts have been reported to explain the various experimental data
with just {\em one}\/ model.  The goal of the present $\vec k$--space
analysis is to provide a transport model that improves on the hitherto
used energy--space descriptions in two important ways. First, we
explictly take into account that {\em in the Au(111) layer no
propagation is allowed along the (111)-direction}. Second, our
scattering dynamics contains no adjustable parameters.

\section{Theoretical model}

Our transport model is based on the semiclassical Boltzmann equation
which is solved in $\vec k$-space by use of the Ensemble--Monte-Carlo
({\sc emc}) technique \cite{Hohenester:93}.

A recent Green--function analysis \cite{deAndres:96} has shown that the
{\sc stm} electrons achieve their bulk Bloch
character, with propagation gaps due to forbidden regions of phase
space, after passing roughly 50 \AA\ within the Au(111) layer, with the
asymptotic form $\sim 1/\cos\theta$ and $\theta\in(20^\circ,70^\circ)$;
this distribution differs appreciably from the hitherto assumed
distribution for an isotropic bandstructure which is concentrated
within a narrow forward cone.
As the mean free path of the injected electrons in
the gold layer is usually larger than 50 \AA, our {\sc emc}
simulations use the asymptotic angular distribution for the input ensemble
of injected {\sc stm}
electrons at the surface. Simulating quasifree electrons ($m_{\rm
eff}=m_o$), we approximately correct for band-structure effects of the
electron
propagation by cutting off the forbidden directions arising from gaps
in the constant-energy surface.  For our case of injection energies
about 1 eV above the Fermi energy, these ''propagation gaps`` form
cones with an opening angle of 10 degrees around the (111) directions
and are included in the scattering dynamics by use of Monte-Carlo
rejection techniques.
The scattering between the
hot electrons and those of the ''cold`` metallic background is treated
via a dynamically (in RPA) screened Coulomb potential, and the
electron-phonon
scattering with an experimentally determined acoustic deformation
potential \cite{Blatt:68}.

Assuming specular transmission/reflection at the Au/Si
interface (via wavefunction matching at a step-like Schottky barrier of
0.8 eV) and either specular or diffuse reflection of backward--running
electrons at the free metal
surface (both types resulting in practically identical simulated {\sc
beem} currents), the boundary scatterings are treated in the
conventional way \cite{Prietsch:95}.

The simulation of each electron is followed in $\vec k$- and $\vec
r$-space and stopped after it has passed the interface or when its
energy has dropped below the top of the barrier. Finally, we assume
negligible current modifications in the semiconductor, which seems
reasonable for the modest electron energies of our present concern.

\section{Results}

Our simulated Au/Si(111) {\sc beem} spectra compare reasonably well with
those in
Fig.1 of Ref.\cite{Bell:96}, with the exception of
the 300\AA\ sample at 77 $K$ where
our simulations completely fail to reproduce the experimental finding
of a drastically altered shape and magnitude \cite{Oxford:97} (see also
symbol $\diamond$ in our Fig.1). We note that for its interpretation Bell
\cite{Bell:96} had proposed a model
based on a narrow forward injection cone, i.e. propagation {\em
inside the forbidden gaps}.

To further inquire into this problematics we compare in Fig.~1, for a
constant tip voltage of $1.2$ V, our theoretical results (using a tunneling
distribution of energetic width 0.5 eV) with various
experimental {\sc beem} data
\cite{Bell:96,Ventrice:96,Schowalter/Lee:91,Lee/Schowalter:92,Manke:95,%
Palm:94,Girardin:96}.
Good agreement with the experimental trends is found, e.g. (i) for
Au/Si(100) the {\sc beem} current at 77 K is {\em always}\/ larger than the
current at
room-temperature, (ii) for almost all experimental data the {\sc
beem} current for Au/Si(100) is larger than the current for Au/Si(111).

We emphasize that a number of important ingredients is still missing in our
transport description, as e.g. phonon--induced backscatterings in the
image--charge potential region of Si
\cite{Ventrice:96}, effects of the non--isotropic band structure on the
scattering dynamics, a possible mismatch of the in--plane $\vec
k$--vector at the interface \cite{Prietsch:95}, more pronounced
anisotropies of the injected electron distribution, and non--ideal
tip--surface geometries \cite{Kalka:95}. Nevertheless, the fact that  our
parameter--free calculations yield qualitative agreement with the (widely
scattered) experimental data gives us confidence in our approach.

\begin{figure}
\epsfig{figure=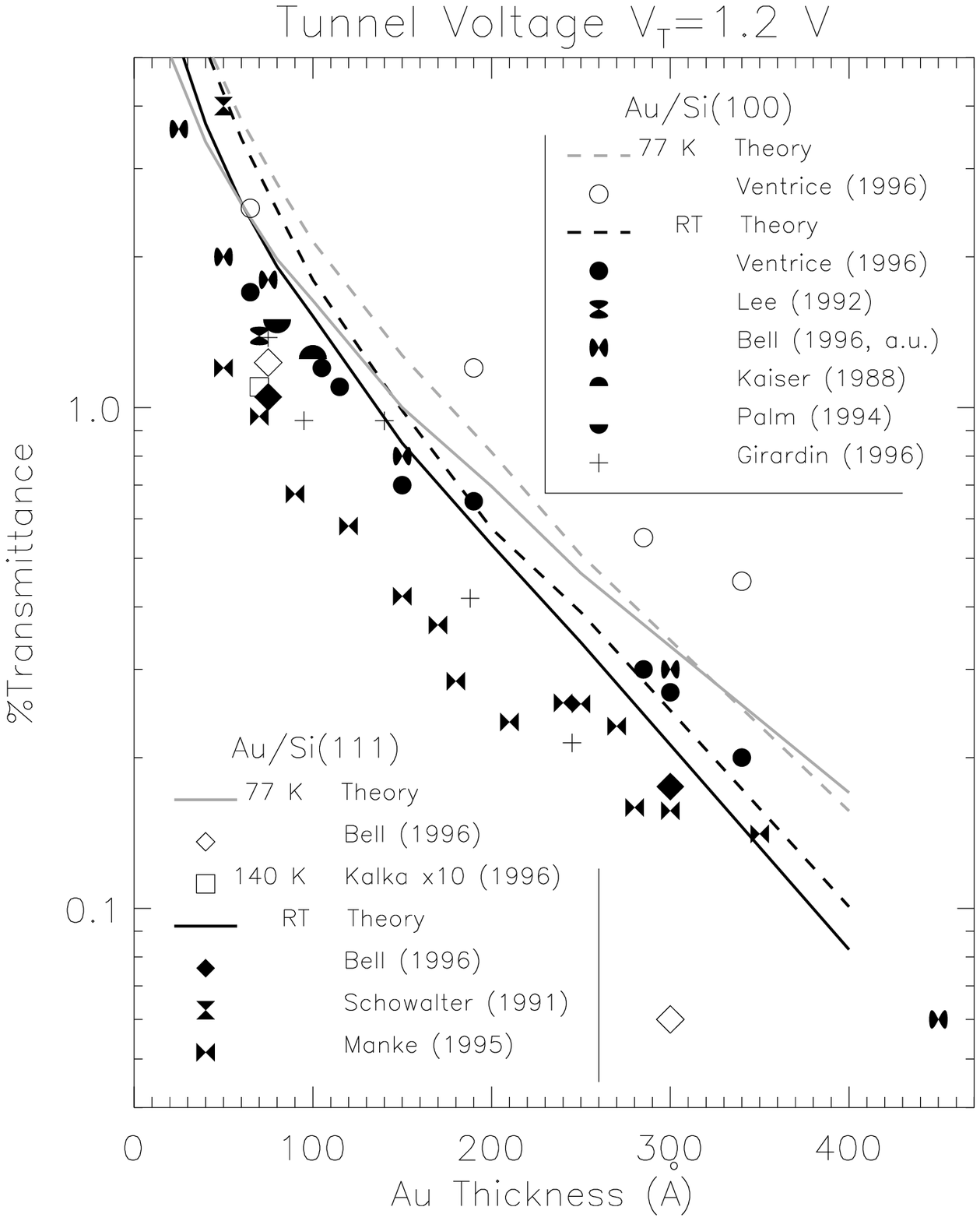,height=8cm,width=8cm}
\caption{Transmittance of the Au(111)/Si interface 
at a tunnel voltage of
1.2 eV as function of the Au--layer thickness.}
\label{fig:elssmx-vs-p.eps}
\end{figure}

\end{document}